\documentstyle[12pt]{article}

\advance\textwidth 2cm
\advance\oddsidemargin -1.0cm

\def\beq{\begin{equation}}
\def\eeq{\end{equation}}
\def\bea{\begin{eqnarray}}
\def\eea{\end{eqnarray}}

\def\al{{\alpha}}
\def\be{{\beta}}
\def\ga{{\gamma}}
\def\la{{\lambda}}

\def\bbeta{\mbox{\boldmath $\beta$}}

\def\brho{\mbox{\boldmath $\rho$}}
\def\bphi{\mbox{\boldmath $\phi$}}
\def\bpi{\mbox{\boldmath $\pi$}}
\def\bpsi{\mbox{\boldmath $\psi$}}
\def\bg{{\mathbf{g}}}
\def\barbg{\bg}
\def\bA{{\bf A}}
\def\bB{{\bf B}}
\def\bD{{\bf D}}
\def\bE{{\bf E}}
\def\bF{{\bf F}}
\def\bH{{\bf H}}
\def\bJ{{\bf J}}
\def\bK{{\bf K}}
\def\bS{{\bf S}}
\def\bfGam{{\mathbf{\Gamma}}}

\def\cC{{\mathcal{C}}}

\def\bGam{{\bar\Gamma}}
\def\bbGam{\mbox{\boldmath $\bGam$}}

\def\d{{\partial}}
\def\dzeroh{{\hat\partial_0}}
\def\grad{{\nabla}}
\def\bgrad{{\bar \nabla}}
\def\bbgrad{{\mathbf{\bar\nabla}}}
\def\Lie{{\pounds}}
\def\Tr{{\mathrm{Tr}}}
\def\tr{{\mathrm{Tr}}}
\def\trK{{\Tr\bK}}

\def\tN{ N \mbox{\kern -0.8em\lower .65ex\hbox{$\scriptscriptstyle\sim$}}}

\def\Boxh{\hat{\mbox{\kern-.0em\lower.3ex\hbox{$\Box$}}}}

\begin{document}

\title{CURVATURE-BASED HYPERBOLIC SYSTEMS\\ FOR GENERAL RELATIVITY}

\author{YVONNE CHOQUET-BRUHAT \\
Gravitation et Cosmologie Relativiste\\ 
t.22-12, Un. Paris VI\\
Paris 75252 France\\
\\
JAMES W. YORK, JR. and ARLEN ANDERSON\\
Department of Physics and Astronomy\\
Univ. North Carolina\\
Chapel Hill NC 27599-3255 USA}

\date{Feb. 6, 1998}

\maketitle
\vspace{-12cm}
\hfill IFP-UNC-523

\hfill TAR-UNC-065

\hfill gr-qc/9802027
\vspace{10cm}
\begin{abstract}
We review curvature-based hyperbolic forms of the evolution part
of the Cauchy problem of General Relativity that we have obtained
recently.  We emphasize first order symmetrizable hyperbolic systems
possessing only physical characteristics.
\end{abstract}

\newpage

\section{Introduction}

Application of general relativity to fully three dimensional problems
in astrophysics and cosmology has provided a driving force motivating
further study and reformulation of Einstein's equations in 3+1 form.
The vital role of the four constraint equations in setting up initial
data for numerical evolution and in carrying out mathematical studies
of gravitational field configurations has long been recognized.  It led
in the 1970's to a general, mathematically rigorous, and useful formulation
of the constraints. (See, for example, the review \cite{cby80}.)
The standard evolution equations for the spatial metric $\bg$ and
the extrinsic curvature $\bK$ were already known from a
straightforward decomposition of the spacetime Ricci tensor (cf.
the cases of zero shift\,\cite{Lic39}, arbitrary shift\,\cite{cb56}, and
synthesis from the spacetime viewpoint and further developments\,\cite{Yor79}).
It was widely believed that these equations (first order in time
derivatives; second order in space derivatives) were adequate for
applications.  However, they do not constitute a hyperbolic system
leading to a proof of causal evolution in local Sobolev spaces of
$\bg$ and $\bK$ into Einsteinian spacetime.  Hence, further study of
evolution equations is essential.

In this paper, we review methods that we have used recently to obtain
hyperbolic systems for the evolution of $(\bg,\bK)$ with only
physical characteristics and which propagate curvature along the
physical light cone.  We emphasize first order symmetrizable hyperbolic
systems.

\section{Einstein-Ricci Hyperbolic System}

What we may call the ``Einstein-Ricci'' (ER) system is a spatially
covariant hyperbolic formulation of the Einstein equations constructed
from first derivatives of the spacetime Ricci tensor.  The ER equations
are obtained from the 3+1 form of the metric
\beq
ds^2 = - N^2 dt^2 + g_{ij} (dx^i + \beta^i dt)(dx^j + \beta^j dt),
\eeq
where $N>0$ is the lapse, $\beta^i$ is the shift, and the properly
Riemannian metric $g_{ij}$ is the
spatial metric of a spacelike slice $\Sigma$.  ($\Sigma$ is
understood to be a generic spacelike leaf of a foliation of a globally 
hyperbolic spacetime: locally, $t$=const.)  It is convenient to work in
the cobasis $\theta^0 = dt$, $\theta^i = dx^i + \beta^i dt$, with
$d\theta^\alpha = -{1\over 2} C^\alpha\mathstrut_{\beta\gamma} \theta^\beta
\wedge \theta^\gamma$ and only $C^i\mathstrut_{0j}= - C^i\mathstrut_{j0}
= \d_j \beta^i \ne 0$.  The connection one-forms 
$\omega^\alpha\mathstrut_\beta = \gamma^\alpha\mathstrut_{\beta\gamma}
\theta^\gamma$ are given by the connection coefficients
\beq
\gamma^\alpha\mathstrut_{\beta\gamma} = \Gamma^\alpha\mathstrut_{\beta\gamma}
+ g^{\alpha \delta} C^\epsilon\mathstrut_{\delta (\beta} g_{\gamma) \epsilon}
-{1\over 2} C^\alpha\mathstrut_{\beta\gamma},
\eeq
where $\bfGam$ denotes a Christoffel symbol.  In particular, the
extrinsic curvature of $\Sigma$ is $K_{ij} \equiv -N \gamma^0\mathstrut_{ij}$,
or, upon evaluation of $\gamma^0\mathstrut_{ij}$,
\beq
\label{dg}
\dzeroh g_{ij} = -2 N K_{ij},
\eeq
where $\dzeroh = \d_t - \Lie_\beta$ evolves $t$-dependent spatial objects
in the direction orthogonal to $\Sigma$, and $\Lie_\beta$ is the Lie
derivative in $\Sigma$ along the shift vector.  The relation (\ref{dg})
serves as the evolution equation for $g_{ij}$.  Letting an overbar
signify spatial objects, we note for future reference that $\bar g_{ij}=
g_{ij}$, $\bar g^{ij} = g^{ij}$, and $\gamma^i\mathstrut_{jk}
=\Gamma^i\mathstrut_{jk} = \bar\Gamma^i\mathstrut_{jk}$.  For
completeness, we also state $\gamma^i\mathstrut_{00} = N \bgrad^i N$,
$\gamma^0\mathstrut_{00} = N^{-1} \dzeroh N$, $\gamma^0\mathstrut_{0i}
=\gamma^0\mathstrut_{i0} = N^{-1} \bgrad_i N$, 
$\gamma^k\mathstrut_{0i} = -N K^k\mathstrut_i$, and
$\gamma^k\mathstrut_{i0} = \gamma^k\mathstrut_{0i} + \d_i \beta^k$.

The 3+1 decomposition of the spacetime Riemann tensor is given, for
example, in Ref.~[4].  Likewise, the Ricci tensor is given
identically by
\bea
\label{Rij}
R_{ij} &\equiv& \bar R_{ij} - N^{-1} \dzeroh K_{ij} + H K_{ij} -
2 K_{ik} K^k\mathstrut_j - N^{-1} \bgrad_i \bgrad_j N, \\
\label{R0i}
R_{0i} &\equiv& N \bgrad^j(H g_{ij} - K_{ij}), \\
\label{R00}
R_{00} &\equiv& N \dzeroh H - N^2 K_{mk} K^{mk} + N \bgrad^k \bgrad_k N,
\eea
where $H\equiv K^k\mathstrut_k \equiv K$.  The heart of the ER 
system\,\cite{cby95}-\cite{aacby97a} is defined by
\beq
\Omega_{ij} \equiv \dzeroh R_{ij} - 2 \bgrad_{(i} R_{j)0}.
\eeq
Upon working out this identity using (\ref{Rij})-(\ref{R00}) and 
substituting the Einstein equations
$R_{\alpha\beta} = \rho_{\alpha \beta} = 8\pi (T_{\alpha \beta} -
{1\over 2} g_{\alpha \beta} T^\gamma\mathstrut_\gamma)$, with
$G=c=1$, we obtain the equation
\beq
\label{BoxK}
-N \Boxh K_{ij} = J_{ij} -S_{ij} + \Omega_{ij},
\eeq
where the physical wave operator for arbitrary shift is
$\Boxh \equiv -N^{-1} \dzeroh N^{-1} \dzeroh + \bgrad^k \bgrad_k$.
We study the {\it vacuum} case $\Omega_{ij}=0$ here.  In general, the value
of $\Omega_{ij}$ is defined by matter fields.

The detailed form of the right hand side of (\ref{BoxK}) can be found
in Refs.~[5]-[7]; the present conventions are those in
Refs.~[6],[7].  Here we point out that $J_{ij}=J_{ij}[\mbox{\rm 2\ in\ $\bg$;
2\ in\ $N$;1\ in\ $\bK$}]$.  (The numbers in the bracket indicate
the highest order derivatives that occur.)

The slicing-dependent term $S_{ij}$ is given by
\beq
S_{ij} = -N^{-1} \bgrad_i \bgrad_j (\dzeroh N + N^2 H).
\eeq
We observe that the term $\bgrad_i \bgrad_j H = g^{mk} \bgrad_i \bgrad_j
K_{mk}$ would spoil the hyperbolicity of (\ref{BoxK}); therefore,
$S_{ij}$ must be set equal to a functional involving fewer than two 
derivatives of $K_{ij}$.  Notice that
\beq
\label{harmonic}
\dzeroh N + N^2 H \equiv -N^3 \gamma^0 \equiv g^{1/2} \dzeroh (g^{-1/2} N)
\equiv g^{1/2} \dzeroh \alpha ,
\eeq
where $g=\det g_{ij} >0$, $\gamma^0 \equiv \mathstrut^{(4)}g^{\alpha\beta}
\gamma^0\mathstrut_{\alpha\beta}$, and $\alpha\equiv g^{-1/2} N$
is a lapse function of weight -1, first introduced in the 
``algebraic gauge''\,\cite{cbr83} and subsequently as a ``proper-time 
gauge''\,\cite{Tei82}
and in connection with the ``new variables'' program for general
relativity in Hamiltonian form\,\cite{Ash88}, where $\alpha = \tN\, $ is
called the ``densitized lapse.''  The algebraic gauge has  a simple
relationship to ``harmonic time slicing,'' $\gamma^0=0$, as we see 
in (\ref{harmonic}).

One most easily deals with $S_{ij}$ by freely specifying $\alpha(t,x)>0$.
(The shift $\beta^i(t,x)$ is also arbitrary.)  A ``gauge
source'' $f(t,x)$ can be used, as observed by
Friedrich (see \cite{Fri96} for references); here, 
$f(t,x)=\dzeroh \log \alpha$. 
Then the first term in
(\ref{harmonic}) yields a general harmonic time slicing equation for
$N$, namely
\beq
\label{harmonic1}
\dzeroh N + N^2 H = N f(t,x),
\eeq
or equivalently,
\beq
\label{harmonic2}
\dzeroh (g^{-1/2} N) = g^{-1/2} N f(t,x).
\eeq
In this case, the system composed of (\ref{harmonic2}) and the 
third-order equation  resulting from combining (\ref{dg}) and
(\ref{BoxK}), whose principal term is $\Box \partial_0 g_{ij}$,
is quasi-diagonal hyperbolic\,\cite{aacby97b} for the unknowns
$\bg$ and $g^{-1/2} N$.  Therefore, the ``third-order'' ER equations
have a unique, well-posed solution in a suitable Sobolev function
space (local in time and in space; global in space can
be obtained as well), given appropriate initial data\,\cite{cby95,Ler52}.

We wish to verify that every solution of the third order ER system,
equivalently $\{$(\ref{dg}), (\ref{BoxK}), (\ref{harmonic2}) or
(\ref{harmonic1})$\}$, is a solution of the Einstein equations given
suitable initial data.
Thus, suppose $\bg$, $\bK$, and $N$ satisfy these equations.  Then,
by using the contracted Bianchi identities 
$\grad^\beta G_{\beta\alpha}\equiv 0$, and these ER equations, it follows 
that $G_{\beta\alpha}=0$ if $\bg$ and $\bK$ satisfy initially the
usual momentum and Hamiltonian constraints 
($\cC_i\equiv -2 N^{-1} R_{0i} =0$, $\cC = 2 G^0\mathstrut_0=0$),
if $\dzeroh K_{ij}$ satisfies $R_{ij}=0$ initially, and if
initial data $\{\alpha>0,\beta^i\}$ or $\{N>0,\beta^i\}$ are
given\,\cite{cby95,aacby95}. Conversely, because every globally
hyperbolic metric solution of Einstein's equations can be given
in a harmonic time slicing\,\cite{cby80}, it follows that all globally
hyperbolic solutions of Einstein's equations can be reached 
by solving this Einstein-Ricci system. 

The ER system as given holds for any spacetime dimension.  For {\it four}
dimensions, it can be written in first-order symmetrizable hyperbolic
(FOSH) form.  To put a system in first-order form, one introduces
auxiliary variables in place of derivatives of the fundamental 
variables and adds additional equations to describe the evolution of
these new variables.  The crux is whether this process terminates.
Generally existence of a higher order wave equation on the fundamental 
variables is necessary to halt the process.  Though we have a wave equation 
for $\bK$, 
presence of second spatial derivatives of $\bg$ and $N$ in $J_{ij}$ prevent the
reduction to first order form from being obvious.  However, the second 
derivatives of $\bg$ occur only in $\bar R_{ij}$ and $\bar R_{ijkl}$ terms; the
latter, in three space dimensions, can be reduced to $\bar R_{ij}$,
which in turn can be eliminated from $J_{ij}$ by using (\ref{Rij}).
First derivatives of $\bg$ are handled by an evolution 
equation\,\cite{cby95,aacby97a}
for the Christoffel symbol $\bbGam$ (inadvertently omitted in [6]).

On the other hand, to handle the second derivatives of $N$,
one first finds a wave equation for $N$. Then, by applying $\bgrad_i$ to it, a
wave equation for $\bgrad_i N$, or $a_i=\bgrad_i \log N$, is obtained.
{}From this point, reduction to first order form is 
straightforward\,\cite{cby95}-\cite{aacby97a}.
[The wave equation for $N$ follows from applying $\dzeroh$ to 
(\ref{harmonic1}) to get an equation with $\dzeroh \dzeroh N$ and 
$\dzeroh H$.  Using (\ref{R00}), we then find a nonlinear wave
equation that can be written
in terms of $\Boxh N$.  Then $\bgrad_i \Boxh N$ gives the wave equation
$\Boxh a_i$. See Refs. [5], [7] for details.]  One then shows
that the first order system is symmetrizable and hyperbolic.  

In this system, the spacetime
metric $(N,g_{ij})$ {\it evolves} at zero speed along the direction
$\dzeroh$ orthogonal to $t$=constant.  Quantities with dimensions of
curvature {\it propagate} at speed 1 (``$c$'') along the physical light
cone.  There are {\it no} unphysical speeds or directions as there are
in many of the FOSH systems that have so far appeared in the 
literature\,\cite{Fri96,FiM72,FrR94}. Of course, we except
the FOSH form of the ER system\,\cite{cby95}-\cite{aacby97a} 
(described here) and the mathematically equivalent, but more transparent,
``Einstein-Bianchi''\,\cite{cby97,acby97} (EB) system presented in the 
next section.
Having only physical characteristics is essential for an ideal
match of physics and mathematics. In practice, it is also
useful: for example,
consider obtaining the gravitational radiation content of a numerical
Cauchy evolution on a finite grid.  The radiation is propagating
at light speed and is either ``extracted''\,\cite{ral} or
``evolved to null infinity''\,\cite{win} based on field values at 
a finite distance from the source.  We know that gravitational
radiation is curvature propagating at light speed and that is
what the ER system describes.  The description is even more
explicit and transparent in the EB system.

A ``fourth order'' ER system also exists\,\cite{aacby96a}.  One forms
the expression
\beq
\dzeroh \Omega_{ij} + \bgrad_i \bgrad_j R_{00}
\eeq
to obtain it.  It has a well-posed Cauchy problem and is hyperbolic
(``non-strict''\,\cite{LeO}) for any $N>0$ and $\beta^i$. It has
been applied to the non-linear perturbative regime of high frequency
wave propagation\,\cite{cb97a,cb97b} and has been used to obtain
an elegant derivation of gauge-invariant perturbation theory for
the Schwarzschild metric\,\cite{aal98}.

\section{Einstein-Bianchi Hyperbolic System}

The ``Einstein-Bianchi'' (EB) system discussed next was given first in
[16] and with energy estimates and full mathematical detail in [17].
It is similar to an analogous system, obtained by H. Friedrich \cite{Fri96},
that is based on the Weyl tensor and is causal but with additional
unphysical characteristics.
Recall that the Riemann tensor satisfies the Bianchi identities
\beq
\label{Bianchi}
\nabla_\al R_{\be\ga,\la\mu} + \nabla_\ga R_{\al\be,\la\mu} +
\nabla_\be R_{\ga\al,\la\mu} \equiv 0,
\eeq
where we use a comma to stress the two separate antisymmetric index pairs
({\it not} to indicate partial differentiation).
These identities imply by contraction and use of the symmetries
of the Riemann tensor
\beq
\label{Bianchi1c}
\nabla_\al R^\al\mathstrut_{\be,\la\mu} + \nabla_\mu R_{\be\la} 
-\nabla_\la R_{\be\mu} \equiv 0,
\eeq
where the Ricci tensor is defined by 
$$R^{\al}\mathstrut_{\be,\al\mu}=R_{\be\mu}.$$
If the Ricci tensor satisfies the Einstein equations
\beq
\label{Einstein}
R_{\al\be} = \rho_{\al\be},
\eeq
then we have
\beq
\label{Bianchi1c_mat}
\nabla_\al R^\al\mathstrut_{\be,\la\mu} = 
\nabla_\la \rho_{\be\mu} - \nabla_\mu \rho_{\be\la} .
\eeq
Equation (\ref{Bianchi}) with $\{\al\be\ga\} = \{ijk\}$ and
(\ref{Bianchi1c_mat}) with $\be=0$ contain only derivatives
of the Riemann tensor tangent to $\Sigma$; hence, we consider
these equations as constraints (``Bianchi constraints'').  We
shall consider the remaining equations in (\ref{Bianchi}) and 
(\ref{Bianchi1c_mat}) as applying to a double two-form 
$A_{\al\be,\la\mu}$, which is simply a spacetime tensor
antisymmetric in its first and last pairs of indices.  We do
{\it not} suppose {\it a priori} a symmetry between the two
pairs of antisymmetric indices.  These ``Bianchi equations'' are 
\beq
\label{Bianchi_eq1}
\nabla_0 A_{hk,\la\mu} + \nabla_k A_{0h,\la\mu} + 
\nabla_h A_{k0,\la\mu} = 0, 
\eeq
\beq
\label{Bianchi_eq2}
\nabla_0 A^0\mathstrut_{i,\la\mu} + \nabla_h A^h\mathstrut_{i,\la\mu}
= \nabla_\la \rho_{i\mu } - \nabla_\mu \rho_{i\la } \equiv
J_{i,\la\mu }, 
\eeq
where the pair $[\la\mu]$ is either $[0j]$ or $[jl]$.  We next
introduce following Bel\,\cite{Bel} two ``electric'' and two ``magnetic'' space
tensors associated with the double two-form $\bA$, in analogy
to the electric and magnetic vectors associated with the electromagnetic
two-form $\bF$.  That is, we define the ``electric'' tensors by
\beq
\label{electric}
E_{ij} \equiv A^0\mathstrut_{i,0j} = -N^{-2} A_{0i,0j}, 
\eeq
$$
D_{ij} \equiv {1\over 4} \eta_{ihk} \eta_{jlm} A^{hk,lm}, 
$$
while the ``magnetic'' tensors are given by
\bea
\label{magnetic}
H_{ij} \equiv {1\over 2} N^{-1} \eta_{ihk} A^{hk}\mathstrut_{,0j}, \\
B_{ji} \equiv {1\over 2} N^{-1} \eta_{ihk} A_{0j,}\mathstrut^{hk}.
\nonumber
\eea
In these formulae, $\eta_{ijk}$ is the volume form of the space metric
$\bg$.  We note that: (1)~If the double two-form $\bA$ is symmetric with 
respect to its two pairs of
antisymmetric indices, then $E_{ij}=E_{ji}$, $D_{ij}=D_{ji}$, and
$H_{ij}=B_{ji}$. (2)~If $\bA$ is a symmetric double two-form such that
$A_{\al\be} \equiv A^{\la}\mathstrut_{\al,\la\be} = c g_{\al\be}$,
then $H_{ij}=H_{ji}=B_{ji}=B_{ij}$ and $E_{ij}=D_{ij}$. 
Property (1) is obvious and (2) follows from the ``Lanczos 
identity.''\,\cite{Lan} (We note that $\eta_{0ijk}=N\eta_{ijk}$ relates
the spacetime and space volume forms in using the Lanczos identity.)

In order to extend the treatment to the non-vacuum case and to avoid 
introducing
unphysical characteristics in the solution of the Bianchi equations, we will
keep as independent unknowns the four tensors $\bE$, $\bD$, $\bB$, and $\bH$,
which will not be regarded necessarily as symmetric.  The symmetries will be
imposed eventually on the initial data and shown to be conserved by
evolution.

We now express the Bianchi equations in terms of the time dependent space
tensors $\bE$, $\bH$, $\bD$, and $\bB$.  
We also express spacetime covariant derivatives $\nabla$ of the spacetime
tensor $\bA$ in terms of space covariant derivatives $\bgrad$ and
time derivatives $\dzeroh$ of $\bE$, $\bH$, $\bD$, and $\bB$ by using
the connection coefficients in 3+1 form as given earlier.

The first Bianchi equation (\ref{Bianchi_eq1}) with $[\la\mu]=[0j]$
has the form
\beq
\label{AcycBianchi}
\nabla_0 A_{hk,0j} + \nabla_k A_{0h,0j} -\nabla_h A_{0k,0j} = 0.
\eeq
A calculation  yields (\ref{AcycBianchi}) in the form
\beq
\label{dzeroH}
\dzeroh (\eta^i\mathstrut_{hk} H_{ij}) + 2 N \bgrad_{[h} E_{k]j} +
(L_1)_{hk,j} = 0, 
\eeq
\bea
\label{L1}
(L_1)_{hk,j} &\equiv&
 N K^l\mathstrut_j \eta^i\mathstrut_{hk} H_{il}
+ 2 (\bgrad_{[h} N) E_{k]j}
+ 2 N \eta^i\mathstrut_{lj} K^l\mathstrut_{[k}B_{h]i} \\
&&\hspace{1cm}
-(\bgrad^l N) \eta^i\mathstrut_{hk} \eta^m\mathstrut_{lj} D_{im}.
\nonumber
\eea
The second Bianchi equation (\ref{Bianchi_eq2}), with $[\la\mu]=[0j]$,
has the form
\beq
\nabla_0 A^0\mathstrut_{i,0j} + \nabla_h A^h\mathstrut_{i,0j} = J_{i,0j},
\eeq
where $\bJ$ is zero in vacuum.  A calculation similar to the one above
yields for the second Bianchi equation
\beq
\label{dzeroE}
\dzeroh E_{ij} - N \eta^{hl}\mathstrut_i \bgrad_h H_{lj} + (L_2)_{ij} =
J_{i,0j},
\eeq
\bea
\label{L2}
(L_2)_{ij} &\equiv& -N(\tr \bK) E_{ij} + N K^k\mathstrut_j E_{ik}
+2 N K_i\mathstrut^k E_{kj} \\
&&\hspace{1cm}  
- (\bgrad_h N) \eta^{hl}\mathstrut_i H_{lj}
+ N K^k\mathstrut_h \eta^{lh}\mathstrut_i \eta^{m}\mathstrut_{kj} D_{lm}
+ (\bgrad^k N) \eta^l\mathstrut_{kj} B_{il}. \nonumber
\eea

The non-principal terms in the first two Bianchi equations  (\ref{dzeroH})
and (\ref{dzeroE}) are linear in $\bE$, $\bD$, $\bB$, and $\bH$, as well
as in the other geometrical elements $N\bK$ and $\bbgrad N$.  The
characteristic matrix of the principal terms is symmetrizable.  The
unknowns $E_{i(j)}$ and $H_{i(j)}$, with fixed $j$ and $i=1,2,3$ appear
only in the equations with given $j$.  The other unknowns appear in 
non-principal terms.  The characteristic matrix is composed of three
blocks around the diagonal, each corresponding to one given $j$.

The $j^{\rm th}$ block of the characteristic matrix in an orthonormal
frame for the space metric $\barbg$, with unknowns listed horizontally
and equations listed vertically, ($j$ is suppressed) is given by
\beq
\label{matrix}
\begin{array}{cc}
\begin{array}{c}

\end{array}
&
\begin{array}{cccccc}
\ \ E_1\ \  & \ \ E_2\ \  & \ \ E_3\ \  & \ \ H_1\ \  & 
\ \ H_2\ \  & \ \ H_3\ \ 
\end{array}
\\
\begin{array}{c}
(\ref{dzeroE})_1 \\
(\ref{dzeroE})_2 \\
(\ref{dzeroE})_3 \\
(\ref{dzeroH})_{23} \\
(\ref{dzeroH})_{31} \\
(\ref{dzeroH})_{12}
\end{array}
&
\left(
\begin{array}{cccccc}
\xi_0 & 0 & 0 & 0 & N \xi_3 & - N \xi_2 \\
0 & \xi_0 & 0 & - N \xi_3 & 0 & N \xi_1 \\
0 & 0 & \xi_0 & N \xi_2 & - N \xi_1 & 0 \\
0 & - N \xi_3 & N \xi_2 & \xi_0 & 0 & 0 \\
N \xi_3 & 0 & - N \xi_1 & 0 & \xi_0 & 0 \\
- N \xi_2 & N \xi_1 & 0 & 0 & 0 & \xi_0 
\end{array}
\right) .
\end{array}
\eeq
This matrix is symmetric and its determinant is the characteristic polynomial 
of the $\bE$, $\bH$ system.  It is given by 
\beq
-N^6 (\xi_0 \xi^0) (\xi_\al \xi^\al)^2.
\eeq
The characteristic matrix is symmetric in an orthonormal space frame 
and the timelike direction defined by $\dzeroh$ has a coefficient
matrix $T_0$ that is positive definite (here $T_0$ is the unit matrix).
Therefore, the first order system is symmetrizable hyperbolic.  We will
not compute the symmetrized form explicitly here.

The second pair of Bianchi equations is obtained from (\ref{Bianchi_eq1})
and (\ref{Bianchi_eq2}) with $[\la\mu]=[lm]$.  We obtain from 
(\ref{Bianchi_eq1})
\beq
\label{dzeroD}
\dzeroh (\eta^i\mathstrut_{hk} \eta^j\mathstrut_{lm} D_{ij} )
+ 2 N \eta^j\mathstrut_{lm} \bgrad_{[k} B_{h]j} +
(L_3)_{hk,lm} = 0,
\eeq
\bea
\label{L3}
(L_3)_{hk,lm} &\equiv& 
2 N \eta^n\mathstrut_{j[m} K^j\mathstrut_{l]} \eta^i\mathstrut_{hk} D_{in} 
+ 2 \eta^j\mathstrut_{lm} (\bgrad_{[k} N) B_{h]j} \\
&&\hspace{1cm} 
+ 2 N K_{l[h} E_{k]m} + 2 N K_{m[k} E_{h]l}
+ 2 H_{i[l} ( \bgrad_{m]} N ) \eta^i\mathstrut_{hk}. \nonumber
\eea
Analogously, from (\ref{Bianchi_eq2}) we obtain
\beq
\label{dzeroB}
\dzeroh (\eta^j\mathstrut_{lm} B_{ij}) 
- N \eta^{kh}\mathstrut_i  \eta^j\mathstrut_{lm} \bgrad_h D_{kj}
+ (L_4)_{i,lm} = -N J_{i,lm},
\eeq
\bea
\label{L4}
(L_4)_{i,lm} &\equiv& 
-N (\tr \bK) \eta^j\mathstrut_{lm} B_{ij}
+ 2 N \eta^h\mathstrut_{j[m} K^j\mathstrut_{l]} B_{ih}
+ 2 N K^h\mathstrut_i \eta^j\mathstrut_{lm} B_{hj} \\
&&\hspace{1cm} 
- (\bgrad_j N) \eta^{hj}\mathstrut_i  \eta^n\mathstrut_{lm} D_{hn}
- 2 N \eta^j\mathstrut_{hi} H_{j[m} K^h\mathstrut_{l]} 
+ 2 E_{i[m} \bgrad_{l]} N
.\nonumber 
\eea

Consider the system (\ref{dzeroD}) and (\ref{dzeroB}) with $[lm]$
fixed.  Then $j$ in $\eta_{jlm}$ is also fixed.  The characteristic
matrix for the $[lm]$ equations, with unknowns $D_{ij}$ and $B_{ij}$,
$j$ fixed, with an orthonormal space frame, is the same as the matrix
(\ref{matrix}).

If the spacetime metric is considered as given, as well as the
sources, the Bianchi equations (\ref{dzeroH}), (\ref{dzeroE}), (\ref{dzeroD}),
and (\ref{dzeroB}) form a linear symmetric hyperbolic system with domain of
dependence determined by the light cone of the spacetime metric.  
The coefficients of the
terms of order zero are $\bbgrad N$ or $N\bK$.  The system is homogeneous
in vacuum (zero sources).

\section{Determination of ($\mathbf{\bar\Gamma,K}$) from Knowledge
of the Bianchi Fields}

We next link the metric and connection to our Bianchi fields.  This
link uses and extends an idea introduced by Friedrich \cite{Fri96} in
his Weyl-tensor construction mentioned above.
We will need the $3+1$ decomposition of the Riemann tensor, which 
is\,\cite{Yor79}
\bea
\label{Rijkl}
R_{ij,kl} &=& \bar R_{ij,kl} + 2 K_{i[k} K_{l]j}, \\
\label{R0ijk}
R_{0i,jk} &=& 2 N \bgrad_{[j} K_{k]i}, \\
\label{R0i0j}
R_{0i,0j} &=& N (\dzeroh K_{ij} + N K_{ik} K^{k}\mathstrut_j +
\bgrad_i \d_j N).
\eea
{}From these formulae one obtains those for the Ricci
curvature given in Sec. 2: (\ref{Rij}), (\ref{R0i}), and
(\ref{R00}), where, in this section, we will {\it not} denote
$K^j\mathstrut_j = \trK$ by $H$.
The identity (\ref{dg})
and the expression for the spatial Christoffel symbols give
\beq
\label{dGam}
\dzeroh \bar\Gamma^h\mathstrut_{ij} \equiv
\bgrad^h(N K_{ij}) - 2 \bgrad_{(i}( N K_{j)}\mathstrut^h).
\eeq
Therefore, from the identity (\ref{R0ijk}), we obtain the
identity
\beq
\label{GamEq}
\dzeroh \bar\Gamma^h\mathstrut_{ij} + N \bgrad^h K_{ij} =
K_{ij} \d^h N - 2 K^h\mathstrut_{(i} \d_{j)} N -
2 R_{0(i,j)}\mathstrut^h.
\eeq
On the other hand, the identities (\ref{R0i0j}) and (\ref{Rij})
imply the identity
\beq
\label{KEq}
\dzeroh K_{ij} + N \bar R_{ij} + \bgrad_j \d_i N \equiv
-2 N R^{0}\mathstrut_{i,0j} - N (\trK ) K_{ij} + N R_{ij}.
\eeq
We obtain equations relating $\bbGam$ and $\bK$ to a
double two-form $\bA$ and matter sources by replacing, in the identities
(\ref{GamEq}) and (\ref{KEq}), $R_{0(i,j)}\mathstrut^h$ by
$(A_{0(i,j)}\mathstrut^h + A^h\mathstrut_{(j,i)0})/2$, 
$R^0\mathstrut_{i,0j}$ by
$(A^0\mathstrut_{i,0j} + A^0\mathstrut_{j,0i})/2$, and the
Ricci tensor of spacetime by a given tensor $\brho$, zero in
vacuum.  The terms involving $\bA$ are then replaced by Bianchi fields
$\bE,\ \bH,\ \bD$ and $\bB$.

The first set of identities (\ref{GamEq}) leads to equations with
principal terms
\beq
\dzeroh \bGam^h\mathstrut_{ij} + N g^{hk} \d_k K_{ij}.
\eeq
To deduce from the second identity (\ref{KEq}) equations which will
form together with the previous ones a symmetric hyperbolic system, we
set, we use algebraic gauge \cite{cbr83}, as in Sec. 2, by setting
\beq
\label{Ndef}
N= g^{1/2} \al 
\eeq
where $\al$ is a given positive scalar density of weight minus one,
a function of $(t,x^i)$.  (Note that the present $\alpha$ is the
$\alpha^{-1}$ of \cite{cby97} and \cite{acby97}.)  The lapse $N$
is now a derived quantity depending on $g^{1/2}$ and on $\alpha$.  
The use of $\alpha$, if $\bbGam$ denotes the
Christoffel symbols of $\barbg$, implies that
\beq
\bGam^h\mathstrut_{ih} = \d_i \log N - \d_i \log \al.
\eeq

The second set of identities (\ref{KEq}) now yields the following
equations, where $N$ denotes $g^{1/2} \al$,
\bea
\label{KEq2}
\dzeroh K_{ij} + N \d_h \bGam^{h}\mathstrut_{ij} &=& N \bigl[
\bGam^m\mathstrut_{ih} \bGam^{h}\mathstrut_{jm} -
(\bGam^h\mathstrut_{ih} + \d_i \log \al) (\bGam^k\mathstrut_{jk}
+\d_j \log \al) \bigr] \\
&&\hspace{-2.5cm} - N (\d_i\d_j \log \al - \bGam^k\mathstrut_{ij}
\d_k \log \al) - N (E_{ij} + E_{ji})- N (\trK ) K_{ij} + N \rho_{ij}. \nonumber
\eea
The first set (\ref{GamEq}) yields
\bea
\label{GamEq2}
\dzeroh \bGam^h\mathstrut_{ij} + N \bgrad^h K_{ij} &=&
N K_{ij} g^{hk} ( \bGam^m\mathstrut_{mk} + \d_k \log \al) \\
&& \hspace{-1cm}
-2 N K^h\mathstrut_{(i}( \bGam^m\mathstrut_{j)m} + \d_{j)} \log \al) 
-N( \eta^k\mathstrut_{(j}\mathstrut^h B_{i) k}
+ H_{k(i} \eta^k\mathstrut_{j)}\mathstrut^h). \nonumber
\eea

We see from the principal parts of (\ref{KEq2}) and (\ref{GamEq2})
that the system obtained for $\bK$ and $\bbGam$ has
a characteristic matrix composed of six blocks around the diagonal,
each block a four-by-four matrix that is symmetrizable hyperbolic
with characteristic polynomial $N^4(\xi_0\xi^0)(\xi_\al \xi^\al)$.
The characteristic matrix in a spatial orthonormal frame has blocks
of the form
\beq
\left(
\begin{array}{cccc}
\xi_0 & N \xi_1 & N \xi_2 & N \xi_3 \\
N \xi_1 & \xi_0 & 0 & 0 \\
N \xi_2 & 0 & \xi_0 & 0 \\
N \xi_3 & 0 & 0 & \xi_0 
\end{array}
\right).
\eeq
\vspace{0.2cm}

\section{Symmetric Hyperbolic System for $(\bE,\bH,\bD,\bB,\barbg,\bK,\bbGam)$}

We denote by $\bS$ the system composed of the equations 
(\ref{dzeroH}), (\ref{dzeroE}), (\ref{dzeroD}), (\ref{dzeroB}),
(\ref{dg}), (\ref{KEq2}), and (\ref{GamEq2}), where the lapse
function $N$ is replaced by $g^{1/2} \al$.  This
system is satisfied by solutions of the Einstein equations whose
shift $\bbeta$, hidden in the operator $\dzeroh$, has
the given arbitrary values and whose lapse has the form
$N=g^{1/2} \al$.  (Clearly, any $N>0$ can be
written in this form.)  
{}From the results of the previous sections, we see that for arbitrary $\al$ 
and $\bbeta$, and
given matter sources $\brho$, the system $\bS$ is a first order
symmetrizable hyperbolic system for the unknowns 
$(\bE,\bH,\bD,\bB,\barbg,\bK,\bbGam)$. 
Note that the various elements $\bE$, $\bH$, $\bD$,
$\bB$, $\barbg$, $\bK$, and $\bbGam$ are considered as independent.
For example, {\it a priori}, we neither know that $\bbGam$ denotes
the Christoffel symbols of $\barbg$ nor that $\bE,\ \bH,\ \bD$,
and $\bB$ are identified with components of the Riemann tensor
of spacetime.

We now consider the {\it vacuum} case.  The original Cauchy data for
the Einstein equations are, with $\bphi$ a properly
Riemannian metric and $\bpsi$ a second rank tensor on an initial spacelike
slice $\Sigma_0$,
\beq
\barbg|_0 = \bphi,\quad \bK|_0 = \bpsi.
\eeq
The tensors $\bphi$ and $\bpsi$ must satisfy the constraints, which
read in vacuum,
\beq
R_{0i} = 0,  
\eeq
\bea
0 = G_{00} &\equiv& R_{00} - {1\over 2} g_{00} R \\
&=& R_{00} + {1\over 2} N^2 g^{\al\be} R_{\al\be} \nonumber \\
&=&{1\over 2} N^2 ( \bar R - K_{ij} K^{ij} + (\trK)^2), \nonumber
\eea
with $\bar R= g^{ij} \bar R_{ij}$.  The initial data given by
$\bphi$ determine the Cauchy data $\bGam^{h}\mathstrut_{ij}|_0$ and
thus $\bar R_{ij,kl}|_0$.  Then, $R_{ij,kl}|_0$
is determined by using also $\bpsi$.  To determine the initial
values of the other components of the unknowns of the
system $\bS$, we use the arbitrarily given data $\al$ and
$\bbeta$.  In particular, we use $N= g^{1/2} \al$
to find $R_{0i,jk}|_0$ and to compute $\bgrad_j\d_i N|_0$
appearing in the identity (\ref{Rij}).  We deduce from
(\ref{Rij}) $\dzeroh K_{ij}|_0$ when $R_{ij}=0$, which enables
$R_{0i,0j}$ to be found from (\ref{R0i0j}).  All of the
components of the Riemann tensor of spacetime are then known
on $\Sigma_0$.  We identify them with the corresponding components of
the double two-form $\bA$ on $\Sigma_0$:  the latter have thus
{\it initially} the same symmetries as the Riemann tensor.  We
find the initial values of $(\bE,\bH,\bD,\bB)$ according to their
definitions in terms of $\bA$.

Detailed existence theorems and proofs that solutions of the EB system,
with initial data given as in the preceding paragraph are found in
[18].  Here we give a less detailed argument, which is nevertheless
completely rigorous.

Let the initial data for the vacuum EB system be given as above.  We
know that our symmetrizable hyperbolic system has a unique 
solution.  Because a solution of Einstein's equations in algebraic
gauge [{\it i.e.}, with given $\alpha(t,x)$]---previously proven to
exist\,\cite{cby95}---together with its connection and Riemann
tensor, satisfies the present EB system while taking the same initial values,
that solution must coincide with the solution of the present EB system
in their common domain of existence.

The FOSH form of the ER system and the EB system are completely
equivalent mathematical systems, with the EB system perhaps having
its unknowns arranged in a more transparent way.  For instance, the
characteristic Bianchi fields with respect to a fixed space direction
that propagate along the light cone are the directionally transverse 
components of the Riemann tensor.  For a detailed discussion of
charateristic fields of the FOSH form of the ER system--also
curvatures--see [7].  These fields for the EB system will be
discussed in more detail elsewhere, as will ``third order'' and
``fourth order'' forms of the EB system, analogous to those of the
ER system.

\section{A Concluding Remark About Constraint Violations}

It is well known that the standard 3+1 equations (\ref{dg})-(\ref{R00}), 
with time derivatives $\dzeroh g_{ij}$ and $\dzeroh K_{ij}$ only,
were reformulated
in terms of closely related variables and written in Hamiltonian
form in far-reaching work by Arnowitt, Deser, and Misner\,\cite{ADM} (ADM)
and by Dirac\,\cite{Dir}.  Indeed, the standard 3+1 equations are often
referred to as the ``ADM equations.''  However, it can be 
shown\,\cite{Fri97,AnY} that when there are violations of the constraints
(as there generically are in numerical work), the standard
3+1 evolution equations and the ADM evolution equations are not
equivalent.  From an analytic standpoint, the standard 3+1 equations
in terms of $(\bg,\bK,R_{ij},R_{0i},G^0\mathstrut_0)$ produce a first order 
symmetric hyperbolic system for the
constraint violations, to which suitable ``energy'' bounds for the 
growth of these violations can be applied, while the actual ADM
equations in terms of $(\bg,\bpi,G_{\mu\nu})$ do not produce a 
hyperbolic system for evolution of constraint violations.  
However, neither the standard 3+1 equations nor the ADM equations
are known to be well-posed, independently of the issue of
constraint conservation.  The ER and EB systems are well-posed,
as are the equations describing conservation of the constraints in these
systems.  This subject is treated in more detail elsewhere\,\cite{AnY98}.

\section*{Acknowledgment} J.W.Y. and A.A. were supported by National 
Science Foundation grants PHY-9413207 and PHY 93-18152/ASC 93-18152 
(ARPA supplemented).

\end{document}